\documentclass[inlimits]{article}[12pt]
\usepackage{mathtext}
\usepackage[T2A]{fontenc}
\usepackage[utf8]{inputenc}
\usepackage[english, russian]{babel}
\usepackage[pdftex]{graphicx}
\usepackage{plain}
\usepackage{amsmath}\usepackage{amssymb}
\usepackage{amscd}
\usepackage{amsthm}

\begin{document}

\title{Применение методов статистического анализа в задаче зрительной рабочей памяти\thanks{Работа выполнена при поддержке РФФИ (проект 18-07-00252)}}
\author{}
\date{}

\maketitle

\author{А.~В. Ерофеева\footnote{Московский государственный университет имени М. В. Ломоносова, факультет вычислительной математики и кибернетики,
yerofeyeva@bk.ru}, Т.~В. Захарова\footnote{Московский государственный университет имени М. В. Ломоносова,
факультет вычислительной математики и кибернетики; Институт проблем информатики Федерального исследовательского
центра «Информатика и управление» Российской академии наук, lsa@cs.msu.ru}}

\vspace{1cm}

\sloppy

{\bf Аннотация:}
Работа посвящена ЭЭГ-исследованию взаимодействия корковых зон мозга, обеспечивающих удержание зрительной информации в рабочей памяти.
Были построены векторные авторегрессионные модели (VAR-модели) сигналов, полученных со связанных с организацией рабочей памяти зон мозга.
Для оценки силы взаимодействия зон по коэффициентам моделей рассчитывалась функция частной направленной когерентности (Partial Directed Coherence, PDC),
основанная на причинности Грейнджера.
Сравнительный анализ для оценки силы внутрикорковых связей был проведен с помощью парного статистического теста Уилкоксона. Найдена зависимость силы связей
от характера выполняемой задачи.

{\bf Ключевые слова:} ЭЭГ; функциональная связность; зрительная рабочая память; PDC; векторная авторегрессионная модель; тест Уилкоксона; причинность Грейнджера

\begin{section}{Введение}

Рабочая память (РП) – многокомпонентная система, несущая функцию временного хранения информации в активном и доступном для анализа состоянии. Рабочая память принимает
участие в мышлении человека, является связующим звеном между системами восприятия, долговременной памяти, а характеристики рабочей памяти определяют скорость и объем
обработки информации. В составе РП выделяют зрительную рабочую память -– подсистему, использующуеся для хранения зрительной информации и для манипулирования ею в ходе
выполнения познавательных заданий.

В настоящее время механизмы рабочей памяти активно изучаются. Хотя была показана ведущая роль префронтальных отделов коры головного мозга в
обеспечении функций рабочей памяти, все чаще исследуются и другие зоны мозга. За последние годы накопилось множество экспериментальных данных, согласно которым
при выполнении людьми заданий, требующих рабочей памяти, активность наблюдается не только в префронтальной, но и в зрительной, теменной, височной коре.
Целью данной работы является изучение связности вышеназванных зон мозга и их взаимодействия при удержании информации в зрительной рабочей памяти.

\end{section}

\begin{section}{Описание эксперимента}

Исследовать функциональную связность мозга предлагается на примере следущего эксперимента. Эксперимент включал в себя две серии испытаний «Наблюдение» и «Рабочая память».
Каждое испытание предполагало наблюдение стимулов – черно-белых изображений прямоугольных решеток с наклоном линий в первой серии, и троек небольших черных кружков на
белом фоне во второй.

В «Наблюдении» для испытуемых не предполагалось когнитивной задачи. В «Рабочей памяти» моделировались ситуации сравнения испытуемым новой и удерживаемой в памяти информации.
В случайном порядке предъявляли пары совпадающих и несовпадающих стимулов. Испытуемый должен был запомнить первый (эталонный) стимул в паре, сравнить его со вторым стимулом
(тестовым) и определить, совпадают ли они.

\end{section}

\begin{section}{Моделирование ЭЭГ векторной авторегрессионной моделью}
Ключевую роль в исследовании играет VAR-моделирование. На основе характеристик построенной модели будет сделан вывод о степени
влияния каналов друг на друга.

Опишем  с помощью VAR-модели порядка \(p\) многоканальный временной ряд ЭЭГ. Пусть дан отрезок оцифрованной \(M\)-канальной ЭЭГ длительностью \(N\) отсчетов\\
\centerline{\(X(t) = [X_1(t), ... , X_M(t)]^T\).}
Тогда \(X(t)\) имеет вид:
\[X(t) = \sum_{i=1}^{p} A(i)X(t-i) + E(t),\]
где \(A(i)\)~-- матрица размерности \(M \times M\), \\
    \(E(t)\)~-- ошибка предсказания, вектор белого шума размера \(M\), \\
    \(p\)~-- порядок модели.

Элемент \(a_{km}(i)\) матрицы \(A(i)\) показывает
какой вклад в текущее значение сигнала \(X_k(t)\) вносит сигнал
\(X_m(t-i)\). Полное влияние канала \(m\) на канал \(k\) определяется
последовательностью коэффицентов \(a_{km}(i)\) при \(i \in \{1, ... , p\}\).
Если коэффиценты равны нулю для всех \(i\), то считаем, что влияние отсутствует.

Следует отметить, что построение VAR-моделей имеет смысл лишь для стационарных сигналов, которым ЭЭГ сигнал в общем случае не является.
Нестационарность ЭЭГ, однако, не исключает применение VAR-моделирования: если найдется промежуток, в пределах которого сигнал стационарен, то построенная на нем VAR-модель
будет давать адекватные результаты. Сигналы со стационарными промежутками называются локально стационарными.

Кроме того, важен выбор порядка VAR-модели, т.е. числа точек в прошлом, которые будут использоваться для предсказания будущего состояния. VAR-модель с малым количеством
задержек аппроксимирует ЭЭГ некачественно, модели высокого порядка более детализированы, но часто ненадежны \cite{kur}.

Для нахождения наилучшего порядка модели может использоваться
критерий Акаике (AIC, Akaike’s Information Criterion). Для фрагмента многоканальной ЭЭГ строится
серия VAR-моделей от \(1\)го до некоторого порядка \(Pm\), заведомо превосходящего
оптимального. Для каждой модели порядка p \(\in \{1, ... ,Pm \}\) определяется
ковариационная матрица остатков \(\Sigma(p)\) и вычисляется значение функции \(AIC\):
\[AIC(p) = N \ln\left( \det \Sigma(p)\right) + 2 p M^2 .\]
Считается, что чем меньше значение, тем лучше модель описывает данные. Однако в качестве оптимального порядка модели рекомендуется выбирать либо точку первого локального
минимума функции \(AIC(p)\), либо в случае монотонного убывания функции ограничивать максимальный порядок, например, величиной \cite{marp}
\begin{equation}
\label{lim}
p_{max} < \frac{3\sqrt{N}}{M}.
\end{equation}
\end{section}

\begin{section}{Расчет частной направленной когерентности}
В последние годы популярность в исследовании связности головного набирают методы, основанные на причинности Грейнджера.
Причинность Грейнджера (Granger causality) -- один из способов исследования влияния временных рядов друг на друга, позволяющий не только находить скрытые взаимодействия
и оценивать их силу, но и определять  направления их действия. Идея этого метода заключается в построении предсказательных моделей, и если модель при учете предыдущих
значений некоторого временного ряда \(X\) точнее предсказывает поведение ряда \(Y\), то считается, что \(X\) влияет на \(Y\) по Грейнджеру.

При исследовании влияния одного канала на другой может потребоваться информация о том  какой вклад вносит каждый ритм. Ритмами называют электрические колебания головного мозга, соответствующие определенному частотному диапазону. Считается, что каждый ритм соответствует
некоторому состоянию мозга и отражает процессы, которые происходят в нем. Характеристики ритма могут быть оценены путем вычисления частотных
характеристик его диапазона.

В рамках векторной авторегрессионной модели в качестве меры влияния по Грейнджеру одного канала на другой на частоте \(f\) может быть
использована Partical Directed Coherence (PDC) – функция частной направленной когерентности.
\[P_{ij}(f) = \frac{\mid A_{ij}(f)\mid}{\sqrt{\sum_{k=1}^M \mid A_{kj}(f)\mid ^2}},\]
где  \(A_{ij}(f)\) -- элемент \(A(f)\),
\[A_{ij}(f) =
\left\{\begin{aligned}
	1 - \sum_{r=1}^p a_{ij}(r)e^{-i2{\pi}fr}, \; & i=j \\
	    \sum_{r=1}^p a_{ij}(r)e^{-i2{\pi}fr}, \; & i \neq j.
\end{aligned}\right.
\]

Элемент \( A_{ij}(f) \) матрицы \(A(f)\) показывает, какой частотной фильтрации
подвергнется сигнал \(X_{j}\) прежде чем стать частью сигнала \(X_{i}\).
Поэтому величину \(P_{ij}(f)\)  можно рассматривать как меру частотного
направленного влияния канала \(j\) на канал \(i\), нормированную на совокупное влияние
\(j\) на все каналы, включая каналы \(i\) и его самого.

\end{section}

\begin{section}{Обработка данных и построение модели}
Из записей ЭЭГ каждого испытуемого исключались фрагменты с физиологическими артефактами и с неверными ответами. Далее выделялись отрезки длиной 900 мс,
соответствующие этапу удержания информации в рабочей памяти, и для каждого слабо стационарного 900 миллисекундного отрезка строилась
векторная авторегрессионная модель.

В качестве каналов модели в полушариях головного мозга были выбраны симметричные
сенсоры, соответствующие затылочному, теменному, височному и лобному отделам коры.
Исходя из расположения электродов, пары были разделены на длинные (пары с
сенсорами лобного отдела коры) и короткие функциональные связи. Всего образовано 12
пар сенсоров: по три пары длинных и коротких связей в каждом полушарии. По данным
этих каналов были рассчитаны коэффициенты авторегрессионных моделей.

Для нахождения оптимального порядка VAR-моделей, из каждого эксперимента случайным
образом выбирались промежутки, на которых строилась функция \(AIC(p)\) для \(p \in \{1, ... , 20\} \).
В каждом эксперименте функции монотонно убывали, и для всех выбранных \(p\) их значения  лежали в
промежутке от 3 до 7. Учитывая верхнюю оценку (\ref{lim}), отрезки для каждого эксперимента моделировались VAR-моделями
15-го порядка.

Коэффициенты моделей использовались для вычисления PDС. Оценки расчитывались для частотного
диапазона от 4 до 30 Гц с шагом в 0.5 Гц для всех стационарных 900 миллисекундных сегментов, а затем
усреднялись по ним для каждой частоты. Описанная выше процедура оценки была необходима для исключения неодинакового
статистического смещения, возникающего при сравнении оценок, полученных для временных
эпох разной длительности \cite{monk}. Полученные значения функции частной направленной когерентности усреднялись в частотных диапазонах
\(\theta\)~- (4-7.5 Гц), \(\alpha\)~- (8-12.5 Гц), \(\beta_1\)~- (13-20.5 Гц),
\(\beta_2\)~- (21-30 Гц).

Так как значения направленной когерентности не являлись нормально распределенными,
статистический анализ их различий для пар сенсоров в задачах
наблюдения и удержания в памяти для каждого из типов стимулов (линии
и паттерны) проводился с использованием парного критерия Уилкоксона с уровнем значимости \(\alpha = 0.05\).

\end{section}

\begin{section}{Результаты}
В работе был проведен сравнительный анализ внутрикорковых связей в задаче зрительной рабочей памяти на
наклон линий и расположение паттернов. Связи оценивались на этапе удержания зрительного стимула и сравнивались с
корковыми связями при простом наблюдении этих же стимулов.

Статистический анализ проведен с помощью парного теста Уилкоксона с уровнем значимости \(\alpha = 0.05\), учитывая поправку Холма-Бонферрони
на множественные сравнения. В рамках каждой серии для каждого полушария проверялось 48 гипотез.

В серии на ориентации линий с учетом поправки Холма-Бонферрони значимые результаты были получены для обоих полушарий. В левом полушарии
значимые различия PDC были найдены для пары сенсоров \(F3 \rightarrow T5\) (нисходящая связь от лобного
сенсора к височному) в \(\alpha, \beta_1\) и \(\beta_2\) диапазонах с pvalue, равными 0.0003, 0.0001,
0.0001 соответственно, а также для пары \(F3 \rightarrow P3\) (нисходящая связь от лобного сенсора
к теменному) в \(\theta\) и \(\beta_2\) диапазонах с pvalue равными 0.0006, 0.001.

В правом полушарии значимое различие PDC было найдено для пары сенсоров
\(F4 \rightarrow T6\) (также нисходящая связь от лобного сенсора к височному) в \(\alpha\) и \(\beta_1\)
диапазонах с pvalue = 0.001 и 0.0007.

В серии на паттерны не было получено значимых результатов с учетом поправки Холма-Бонферрони как для левого, так и для правого полушарий.

\end{section}

\begin{section}{Заключение}
В данной работе был применен метод аппроксимирования ЭЭГ авторегрессионной моделью.
С помощью функции частной направленной когерентности были определены силы влияния каналов ЭЭГ друг на друга.
Использование теста Уилкoксона позволило оценить силу статистической связи между каналами ЭЭГ.
В задаче зрительной рабочей памяти на ориентации линий найдены значимые различия в уровне нисходящей связи от лобной к височной области в обоих полушариях.
Также в задаче на ориентации линий в левом полушарии найдены значимые различия в уровне нисходящей связи от лобной к теменной области.
Таким образом, в работе показана зависимость силы длинных связей от характера выполняемой задачи.
Полученные результаты могут быть полезны в изучении болезней, для которых характерны различные нарушения рабочей памяти.
\end{section}

\newpage

\begin{center}

{\bf Application of statistical analysis to working memory problem}

\medskip

A. V. Erofeeva, T. V. Zakharova

\end{center}

\bigskip

{\bf Abstract:}
This article is devoted to EEG studying of connectivity cortical areas involved in keeping vision information in working memory.  VAR-modeling was used for describing signals got from connected with working memory brain zones. Brain connections were estimated by based in Granger Causality Partial Directed Coherence (PDC) and then compared by Wilcoxon signed-rank test.  In paper connection intensity dependence on executing task was found.

\smallskip


\bigskip

\bigskip

{\bf References}

\begin{enumerate}
\item Thomas B. Christophel, P. Christiaan Klink, Bernhard Spitzer,
Pieter R. Roelfsema, John-Dylan Haynes
\emph{The Distributed Nature of Working Memory.} Trends in Cognitive Sciences, 2017
\item Velichkovsky B.B., Kozlovsky S.A.
\emph{Human working memory: fundamental researches and practical applications.}
\item Utochkin I.S., Yurevich M.A., Bulatova M.E.
\emph{Vision working memory: methods, researches, theories}
\item Mingzhou Ding, Steven L. Bressler, Weiming Yang, Hualou Liang
\emph{Short-window spectral analysis of cortical event-related potentials by adaptive multivariate autoregressive modeling:
data preprocessing, model validation, and variability assessment.} Biological Cybernetics, 2000
\item Kurgansky A.V.
\emph{Some Methodological Issues of Studying Cortico-Cortical Functional Connectivity with Vector Autoregressive Model of Multichannel EEG}, 2010
\item Marple
\emph{Digital Spectral Analysis}
\item Alexey M. Ivanitsky, Andrey R. Nikolaev, George A. Ivanitsky
\emph{Electroencephalography.} Modern Techniques in Neuroscience Research
\item Luiz A. Baccala, Koichi Sameshima
\emph{Partial directed coherence: a new concept in neural structure determination.} Biological Cybernetics, 2001
\item A Schlogl
\emph{The Electroencephalogram and the Adaptive Autoregressive Model: Theory and Applications.}
\end{enumerate}

\end{document}